# Scoping review of methodology for aiding generalisability and transportability of clinical prediction models

Kritchavat Ploddi[1], Matthew Sperrin[1], Glen P. Martin[1], Maurice M. O'Connell[1]

[1] Division of Informatics, Imaging & Data Sciences, Faculty of Biology, Medicine and Health, University of Manchester, Manchester, United Kingdom

**Abstract**
Generalisability and transportability of clinical prediction models (CPMs) refer to their ability to maintain predictive performance when applied to new populations. While CPMs may show good generalisability or transportability to a specific new population, it is rare for a CPM to be developed using methods that prioritise good generalisability or transportability. There is an emerging literature of such techniques; therefore, this scoping review aims to summarise the main methodological approaches, assumptions, advantages, disadvantages and future development of methodology aiding the generalisability/transportability. Relevant articles were systematically searched from MEDLINE, Embase, medRxiv, arxiv databases until September 2023 using a predefined set of search terms. Extracted information included methodology description, assumptions, applied examples, advantages and disadvantages. The searches found 1,761 articles; 172 were retained for full text screening; 18 were finally included. We categorised the methodologies according to whether they were data-driven or knowledge-driven, and whether are specifically tailored for target population. Data-driven approaches range from data augmentation to ensemble methods and density ratio weighting, while knowledge-driven strategies rely on causal methodology. Future research could focus on comparison of such methodologies on simulated and real datasets to identify their strengths specific applicability, as well as synthesising these approaches for enhancing their practical usefulness.

**Introduction**

Generalisability and transportability collectively refer to the extent to which results of research can be applied to settings and populations other than those in which the research was originally conducted. These include an application of the results to different geographical locations, times, or environments. In clinical prediction [1, 2], these terms refer to the ability of a clinical prediction model (CPM) to maintain predictive performance when applied to new settings and populations [3, 4]. Generalisability of a CPM refers to its ability to maintain predictive performance in target populations (the populations in which the model is intended to be applied) which, in general, distributionally overlap with the data the model was developed on; transportability of a CPM refers to its ability to maintain predictive performance in a target population that is separate to the data the model was developed on (Figure 1). Both generalisability and transportability are typically assessed by conducting external validation of a CPM in the target populations [5, 6].

CPMs are usually developed to minimise predictive error in the training data, using methods such as penalised likelihood to mitigate overfitting. This does not guarantee that the models are generalisable or transportable; that is, the predictive error may not be minimised in a target population or setting that differs from the development population. This is often attributed to changes in the distribution of data (such as predictors, outcomes, or predictor-outcome relationships) between training and target population/setting: so-called data distribution shift [7, 8]. The shift could be due to changes in technology (e.g., changes in coding definitions by different EHR platforms used in different hospitals[9]), population and/or setting (e.g., taking a model developed in hospitalised patients to a community population), or behaviour (e.g., adoption of an effective CPM may result in better treatment outcomes and this changes may gradually reduce model performance [9, 10]).

A common way in which poor transportability/generalisability of a CPM is handled is by applying updating methods to tailor the model to the new target population/setting [11, 12]. These methods include updating the regression coefficients according to the new data [13, 14], combining multiple similar models with meta-analysis techniques [4, 15, 16], and continuously updating the models [17, 18]. However, if the generalisability and transportability of a CPM could be improved during the initial development phase, it would readily work in new target populations without any adjustment, thus reducing the need for further updates and the resources required for such updates.

There is emerging literature around methods that seek to develop CPMs with improved generalisability/ transportability, including using causal methods such as a selection diagram or a directed acyclic graph [19, 20]. However, these methods are rarely used in practice. One reason for this might be a lack of awareness of these methods and a current lack of a summary of them for a broad audience.

As such, the aim of this scoping review was to identify methodology that develops clinical prediction models such that they have good generalisability and/or transportability properties. We aim to summarise the main methodological approaches, along with reported assumptions, reported advantages and reported disadvantages. In addition, we will identify any gaps and future development required in methodology for developing generalisable and transportable CPMs.

**Methodology**

**Definitions**

The *development population* is the population from which the data used to develop the CPM was (randomly) sampled. The *target population* is the population in which we are interested in applying the CPM. *Generalisability* refers to maintenance of predictive performance for a target population that contains the development population as a non-random subset. T*ransportability* is maintenance of predictive performance for a target population that is distinct (non-overlapping) from the development population. (See Figure 1)

For example, imagine one develops a CPM to predict cardiovascular disease in adults who are hospitalised in Manchester, using data on adult patients admitted to Manchester hospitals. Imagine we then wish to explore the use of this model in predicting cardiovascular disease in adults admitted to any hospital in the UK. Since Manchester hospitals are a non-random subset of all UK hospitals, then this is assessing the CPM's generalisability. In contrast, if we wished instead to explore the use of this model in predicting cardiovascular risk in patients who visit primary care services, then this would be evaluating the transportability of the model (since such patients are represented by a different population from the hospitalised patients).

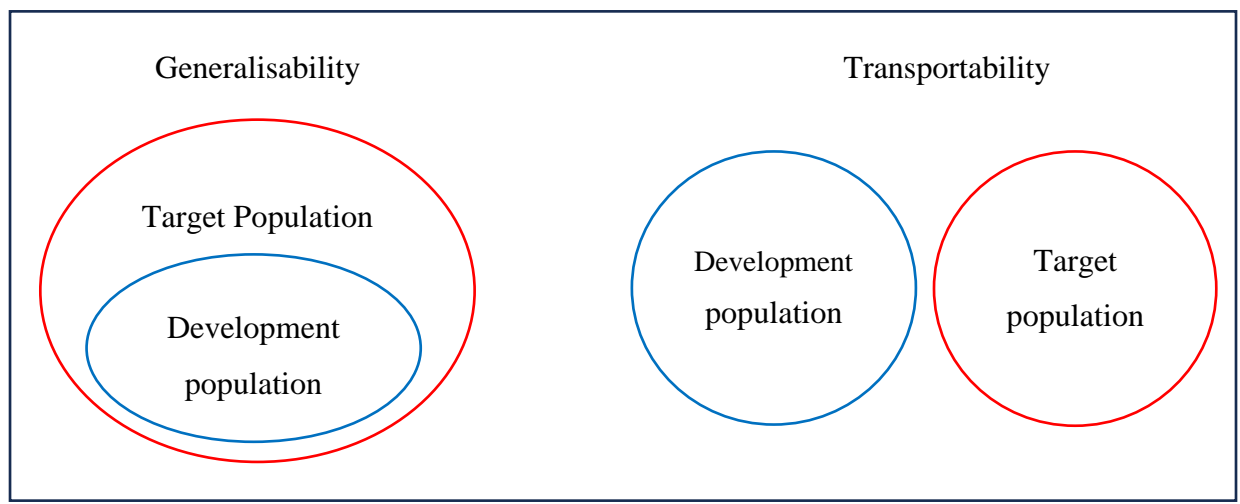

Figure 1. Diagram showing a relationship of the development and target populations: (left) generalisability, (right) transportability.

Scope of review

This scoping review included any methodology that contributes to the development of CPMs that have improved generalisability/transportability across different target populations. The review did not include methodology that takes developed models and make an adjustment to improve generalisability/transportability, such as model updating techniques, for which there are existing reviews [11, 12].

Inclusion/Exclusion criteria

Inclusion

- Papers proposing any method that aims to aid the generalisability or transportability of a CPM, at the time of model development
- Applied papers reporting the development of a CPM, by using specific methods intended to aid the generalisability or transportability of said model

Exclusion

- Non-English papers
- Letters, comments, editorials, and conference abstracts with no information to allow assessment of proposed methods
- Papers that lack sufficient information on how the method improves generalisability or transportability
- Papers only considering an external validation of an existing CPM
- Papers considering model updating, transfer learning or other adjustment methods to improve generalisability or transportability after the initial model has been developed
- Papers without prediction using structured data (i.e data in a tabular format where each column represents predictors or an outcome of interest and each row represent each observation)
- Papers considering out-of-distribution uncertainty in prediction.

We aim to find the methodology that deterministically improves generalisability/transportability rather than reducing the predictive uncertainty so that out-of-distribution uncertainty papers were excluded. However, we acknowledged that the predictive uncertainty significantly influence decision making when applying the models in completely new environments.

Search Strategy

Defining search keywords a priori for methodological papers can be challenging, because of inconsistent methodological terms used by different researchers without standardized terminology [21]. We therefore defined an initial set of search terms using the titles, abstracts and keywords of the following papers related to the generalisability of CPMs that we were aware of prior to conducting this review:

- Bellamy, D., Hernán, M. A., & Beam, A. (2022). A structural characterization of shortcut features for prediction. European Journal of Epidemiology, 37(6), 563–568. [22]
- Piccininni, M., Konigorski, S., Rohmann, J. L., & Kurth, T. (2020). Directed acyclic graphs and causal thinking in clinical risk prediction modeling. BMC Medical Research Methodology, 20(1), 179–179. [23]
- Subbaswamy, A., Schulam, P., & Saria, S. (2018). Preventing Failures Due to Dataset Shift: Learning Predictive Models That Transport. [19]
- Fehr, J., Piccininni, M., Kurth, T., & Konigorski, S. (2022). A causal framework for assessing the transportability of clinical prediction models (p. 2022.03.01.22271617). medRxiv. [20]
- Jong, V. M. T., Moons, K. G. M., Eijkemans, M. J. C., Riley, R. D., & Debray, T. P. A. (2021). Developing more generalizable prediction models from pooled studies and large clustered data sets. Statistics in Medicine, 40(15), 3533–3559. [24]

Specifically, from these papers, we identified the following search terms:

- generalisab*
- generalizab*
- transportab*
- "dataset shift"

We searched for papers having titles including any of these terms (generalis* OR generaliz* OR transportab* OR "dataset shift") to find research mainly focusing on generalisability or transportability. To further filter the results for CPMs, we used search terms proposed by Geersing et al. [25] for identifying prediction modelling studies:

1. (Validat* OR Predict*.ti. OR Rule*) OR (Predict* AND (Outcome* OR Risk* OR Model*)) OR ((History OR Variable* OR Criteria OR Scor* OR Characteristic* OR Finding* OR Factor*) AND (Predict* OR Model* OR Decision* OR Identif* OR Prognos*)) OR (Decision* AND (Model* OR Clinical* OR Logistic Models)) OR (Prognostic AND (History OR Variable* OR Criteria OR Scor* OR Characteristic* OR Finding* OR Factor* OR Model*))
2. Stratification OR ROC Curve OR Discrimination OR Discriminate OR c-statistic OR c statistic OR Area under the curve OR AUC OR Calibration OR Indices OR Algorithm OR Multivariable

Articles were searched based on titles containing terms related to generalisability AND terms in (1) OR (2), i.e. ( generalisability terms ) AND ( CPM term (1) OR CPM term (2) ). The aforementioned five papers were used as a validation of this final search (i.e., we ensured the above five papers were identified from our final search string).

Articles from the arXiv database were exempt from the application of the Geersing et al. criteria, as papers found using generalisability terms were relatively fewer than those from other databases and they typically include recent methodology applying in other fields and having the potential for clinical prediction..

Information sources

The primary author applied the above search terms to MEDLINE, Embase, merRxiv and arXiv, and performed abstract screening and subsequent full-text screening on the filtered articles according to the inclusion/exclusion criteria . We used Ovid Medical Research Platform to implement search strategies for MEDLINE, Embase and medRxiv databases. MedRxiv and arxiv are pre-print databases and are also included as many recent methodological papers are not yet published in peer-review journals.

Snowballing (Citation search)

To compliment the database search, we further applied the inclusion/exclusion criteria to papers which cite, or were cited by, the papers identified from the search (after full-text screen). In addition, we were also aware that other researchers have published titles relating to generalisability in general or other settings but not directly aiding the generalisability of a CPM, such as methods to generalise RCT results to observational data. We did not build search terms based on these but rather did citation searching based on these studies:

- Colnet, B., Mayer, I., Chen, G., Dieng, A., Li, R., Varoquaux, G., Vert, J.P., Josse, J., & Yang, S. (2020). Causal inference methods for combining randomized trials and observational studies: A review. https://doi.org/10.48550/arXiv.2011.08047
- Ling, A. Y., Montez-Rath, M. E., Carita, P., Chandross, K. J., Lucats, L., Meng, Z., Sebastien, B., Kapphahn, K., & Desai, M. (n.d.). An Overview of Current Methods for Real-World Applications to Generalize or Transport Clinical Trial Findings to Target Populations of Interest. Epidemiology, 10.1097/EDE.0000000000001633. https://doi.org/10.1097/EDE.0000000000001633
- Kallus, N., Puli, A. M., & Shalit, U. (2018). Removing Hidden Confounding by Experimental Grounding. Advances in Neural Information Processing Systems, 31. https://dl.acm.org/doi/pdf/10.5555/3327546.3327747
- Dahabreh, I. J., & Hernán, M. A. (2019). Extending inferences from a randomized trial to a target population. European Journal of Epidemiology, 34(8), 719–722. https://doi.org/10.1007/s10654-019-00533-2

Data extraction

Once the papers were screened (both database and snowballing), the primary author (KP) extracted a pre-defined set of information from the included papers. The following information was extracted:

- Title, authors, year of publication.
- Whether generalisability, transportability or both, based on the definition of the target population.
- Whether the methodology is tailored for a specific target population
- Methodology, including the family of proposing methodology (e.g., causal inference), description of the methodology, underlying reported assumptions.
- Any application of the proposed methodology, including in medical and other areas.
- Reported advantages, including reported key strengths of the methodology. This potentially includes sound methodology with explicit examples.
- Reported disadvantages, including reported limitations, challenges or knowledge gaps of the methodology.

**Results**

The database searches identified 1,761 articles. Of these, 172 were retained for full text screening. Finally, 13 articles were eligible for inclusion into the study. Through citation searching, four additional papers were identified and included. One paper from the validation set [22] was not found by the search strategy, but nevertheless it was included in the review, giving the final 18 papers eligible for the scoping review (Figure 2).

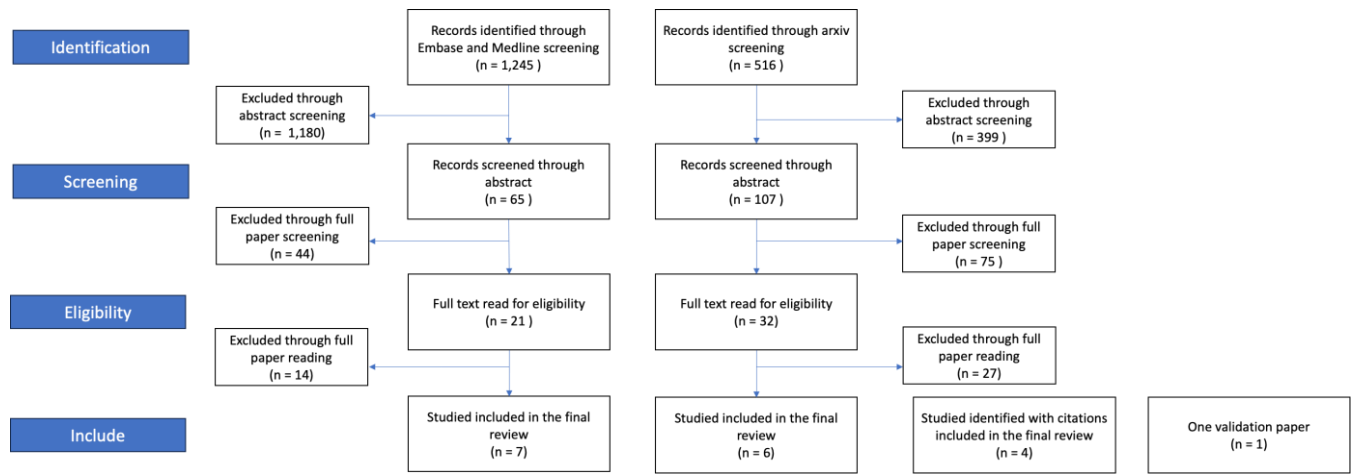

Figure 2. PRISMA (Preferred Reporting Items for Systematic Reviews and Meta-Analysis) flow diagram showing the various stages through the scoping review

The scoping review has identified several methodologies to aid the generalisability and transportability of clinical prediction models. To aid in summarising, we have categorised them into: (1) data-driven methodologies that enhance the robustness of prediction models [23, 24, 26–35], and (2) knowledge-driven methodologies that integrate causal reasoning [19, 20, 22, 23, 30, 36–38], (Table 1 and Supplementary table 1). Some methodologies were grouped into both domains [23, 30]. The data-driven approach aims to enhance model robustness by adopting methodologies primarily informed by data. In contrast, knowledge-driven methods utilise existing domain knowledge about causal relationships between variables to map this understanding into mathematical equations and statistical distributions prior to the application of empirical data.

In addition, we classified methodologies based on whether they are tailored for a specific target population. We found that methodologies that additionally take data from the target population as part of development data primarily aim to develop prediction models tailored for that specific population, while those that do not take data from a target population aim to develop models likely to generalise across various (unstated) target populations [39]. Both classifications are shown in Table 1

|   | Not tailored for target population | Tailored for target population |
|---|---|---|

| Data-driven | 1. Data augmentation [24, 33]<br>2. Ensemble modelling [27] | 1. Density ratio/importance weighting [29]<br>2. Adversarial validation [35]<br>3. Measurement error comparison [29]26/11/2024 12:10:00 |
|---|---|---|
| Knowledge-driven | 1. Selection diagram [19, 37]<br>2. Graph surgery estimator [19]<br>3. Shortcut predictor [22]<br>4. Distributionally robust model [30, 31]<br>5. Minimising worst-case risk [30, 31]<br>6. Optimal stable and immutable sets [30, 31]<br>7. Markov blanket, Parent-child nodes [20, 23, 38]26/11/2024 12:10:00 | Causality-aware prediction by counterfactual approach [36] |

Table 1: Classification of the methodology aiding generalisability and transportability of clinical prediction models

**Data-driven methodologies that are not tailored for target population**

This group of methods involves developing a prediction model using heterogeneous data [24, 27, 33], which can be collected, for example, from multiple centres or locations across different geographical regions. Exposing the model to heterogeneous development data will dilute location- or cluster-specific patterns and this can lead to a more generalisable model [33]. Additionally, the model's transportability could be higher if this heterogeneous development data partly represents the target population in respect to covariate and/or outcome distributions. Some of the methods also enhance the data diversity by augmenting the training data, either through analysing clustered data [24] or by embracing cohort heterogeneity [33]. The strategies improve generalisability by capturing more general patterns of the development data and less subgroup-specific patterns.

Data heterogeneity

de Jong et al. extended the Internal-External Cross-Validation (IECV) technique to develop models that generalise well [24]. IECV non-randomly clusters the training data, ensuring that the selection introduces a level of heterogeneity that mirrors real-world scenarios. Stepwise selection is then performed, with the optimality criteria including a generalisability penalty, thus identifying an optimal prediction model that demonstrates best performance on unseen, heterogeneous data, which potentially results in a better generalisable model. This also can be seen as finding a parsimonious model on a heterogenous data.

Similarly to de Jong et al., Reps (2022) applied ensemble techniques with the objective of improving the model's generalisability to new populations[27]. They trained an L1-regularised logistic regression model on five distinct healthcare databases, which reflect the heterogeneity on the development populations. These base learners were then combined – ensembled – using various techniques, such as weighted and stacked ensembles. The ensembled models are learned from different base learners and different databases so that they can generalise to the new data. Additionally, the authors used a 'leave-one-database-out' approach, where one of the five databases was held out to validate the performance of the ensemble models trained on the remaining four[27], and all held-out validation process was repeated until all databases were evaluated.. The models identified from a 'leave-one-database-out' are expected to have the highest performance on the held-out database, as they were specifically tuned to this database. This resembles cross-validation, which is used to identify any model that is least overfit to the development data.  However, the authors found that while the method improved discriminative performance on a new unseen database, it still had poor calibration (i.e. poor calibration-in-the-large and calibration slope) on the new database [27]. This might reflect that this approach still does not find the optimal prediction model that works best in the new unseen data for all performance metrics.

**Data-driven methodologies that are tailored for target population**

These are methods that additionally take data from target population as part of development data. By exposing parts of the target population during initial model development, the models are more likely to be specifically tailored for that target population. The additional data from the target population could be only covariate data or with an addition of outcome data. This process mainly improves transportability by tailoring the model to the target population.

The first methods involve adjusting the characteristics of individuals in the development data to resemble those in the target data in terms of predictors or covariates by assigning weights to individuals in the development population based on their similarity to those in the target population [26, 28, 32, 34]. These weights are usually represented by the density ratio, which, although calculated slightly differently across various articles, retains a similar underlying concept: making the development population more representative of the target population.

Using the density ratio for weighted maximum likelihood minimises the Kullback-Leibler divergence between the estimated and true conditional density of the outcome given covariate. This results in the prediction models that are specifically tailored for the target population, which improve the transportability [32]. Mathematically, the density ratio, $r(x)$ is the ratio of the probability distribution of covariate between target and development population,

$$r(x) = \frac{P_{target}(x)}{P_{source}(x)} = \frac{P(x|s=0)}{P(x|s=1)},$$

where $P_s(x)$ is the probability distribution of covariate $x$ for the population $s$, for which $s=0$ indicates target population and $s=1$ indicates development population. The ratio can be directly calculated using Kernel Mean Matching (KMM) and Relative Unconstrained Least-Squares Importance Fitting (RuLSIF). Furthermore, the ratio can be indirectly estimated by training a classifier to predict the odds of being sampled from the target population,

$$\frac{P(s=0|x)}{1 - P(s=0|x)} = \frac{P(s=0|x)}{P(s=1|x)}.$$

This indirect estimation were suggested by Gao et al. (2021) [26] and Steingrimsson et al. (2022) [32]. Gao et al. fitted L1- and L2-penalised logistic regression, and random forest to predict how likely each observation having covariate $x$ is being sampled from target population, $s=0$. Note that this estimation requires covariate data from the target population but does not require outcome data. They then fit clinical prediction models by weighted likelihood estimation using these aforementioned odds. Steingrimsson et al. also estimated the density ratio as the ratio of the probability of sampling from the target population, $P(s=0|x)$, to the probability of sampling from the development population $P(s=1|x)$, as shown on the right-hand side of the above equation. They termed this the inverse of the odds of being from the development population. The inverse odds is used for developing prediction models with weighted logistic regression in a manner similar to that of Gao et al. Furthermore, in non-nested designs, in which the development and target populations are completely different and sampling proportions from each populations are unknown unless entire observations are obtained, the inverse odds is not identifiable. However, Steingrimsson et al. demonstrate that the inverse odds is proportional to the inverse odds in the training set,

$$\frac{P(s=0|x)}{P(s=1|x)} \propto \frac{P(s=0, x, D_{train}=1)}{P(s=1, x, D_{train}=1)},$$

under the assumptions where (1) the outcome is conditionally independent of the population given the covariates, i.e. no covariate shift, $(P(y|x,s=0) = P(y|x,s=1))$, (2) the probability of covariate patterns that exists in the target population is not zero in the development population, i.e. overlap/positivity, and (3) the study is non-nested design, where the development and target data are separated sampling from different populations – i.e. transportability. By making these assumptions, especially the last assumption, the methodology aid toward improving the transportability of a clinical prediction model.

Lastly, Dockès (2021) defined importance weights, which are the ratio of the joint probability distribution of covariates and outcome in the target over the development data [28]. This is equal to the density ratio, $r(x)$. These importance weights were used to modify the training errors or loss functions. Dockès also noted that when the importance weights are applied with Bayes' rule in cases of covariate shift exhibiting selection bias, the inverse odds similar to those proposed by the previous researchers is obtained.

The second method was proposed by Qian et al. who introduced the concept of adversarial validation to enhance the transportability of credit score prediction models [35]. This method involves building a classifier to predict the probability of being from the target population on the development data, $P(s = 0 \mid X, s = 1)$. The predicted probabilities are used during an internal validation to find the model that best transports to the target population by either selecting a subset of an internal validation dataset with the highest probabilities, or using the probabilities as sample weights during fitting the models. By modifying internal validation datasets toward target population, a developed model is tailored for the target population, hence improved transportability. Nevertheless, the application of this methodology in clinical prediction models has been limited.

Finally, Pajouheshnia et al. (2019) addressed transportability from a different point of view [29]. They argued that apart from dataset shift, transportability issues could also arise from variations in measurement error patterns between development and target populations, even in the absence of dataset shift. They formally demonstrated that disparate measurement error distributions across populations can lead to suboptimal model performance in the target population. The authors suggested that ensuring consistent measurement processes for predictors in both populations could enhance the transportability of a prediction model and mitigate the risk of performance degradation. This method does not directly improve generalisability and transportability per se, but rather provide the guidance to do so, i.e. reducing measurement error. This is similar to imputing missing data on the development data in order to make the data distribution shift smaller.

**Knowledge-driven methodology that are not tailored for target population**

Fewer studies (7 of 19) focused on knowledge-driven methodologies, 6 of which are not tailored for target population. Here, the methodologies leverage external knowledge about causal relationships, for example, biological processes. Take Acute Myocardial Infarction (AMI) as an example. It is essentially heart muscle death caused by a sudden reduction in blood flow, often due to a ruptured plaque in a blood vessel. These core biological processes tend to be consistent across different settings. In contrast, how these processes are observed can vary. For instance, the pattern of testing for plaque or other causes of reduced blood flow might differ by location. By incorporating causal relationships, these models can be more generalisable. They focus on the underlying biological processes (e.g., AMI) while adjusting for local variations in how those processes manifest (e.g., plaque testing patterns) [40].

These studies built on the structural causal model's framework, which represents causal relationships among predictors and outcomes in forms of a causal diagram – a directed acyclic graph (DAG), and a set of mathematical equations describing such relatmanchionships. Pearl and Bareinboim [37] formalised transportability formulas for causal effects, noting that an identification of a given statistical relation, such as the conditional probability of Y given X, in the target population could be derived from a slightly modified version of the formula.

For example, finding the probability of AMI in the target population, given the probability is estimated from the development population and having a set of predictors that have similar causal relationships (coronary artery plaque size) in both development and target populations – i.e., the probability of AMI given the plaque is independent of being selected from the populations. This independence is called S-admissibility. Specifically, the probability of AMI in the target population can be estimated by standardising the corresponding probability in the development population over a set of covariates that satisfies S-admissibility conditions, assuming the marginal probability of those covariates in the target population is known. For example, $P(Y = 1|X = x, S = 0) = \sum_Z P(Y = 1|X = x, S = 1, Z = z)P(Z = z)$, where P(Y=1) is the probability of AMI, Z is the predictors satisfying S-admissibility (for example, plaque size), X is a set of other predictors, and S is whether being selected from development (s=0) or target (s=1) populations.

This formula can be further extended to a selection diagram (figure 3). It is a DAG having at least one selection variable, which is a variable that determines whether an individual belongs to the development or target population. A selection variable can affect or can be affected by other variables. S-admissibility of a set of predictors is satisfied if the outcome of interest is independent of the selection variables, conditional on the predictors. For example, suppose that we have developed a clinical prediction model using smoking and genetic factors as predictors. However, in the target population both predictors are different in distribution (figure 3). The set of predictors containing smoking and genetic factors are s-admissible for lung cancers if the prevalence of lung cancers are equal between the development and target populations given the genetic factors and smoking. That is, there are no other causes of lung cancers that also differ in both populations. In this setting, prediction of lung cancers using smoking status and genetic factors can be transported to the target population.

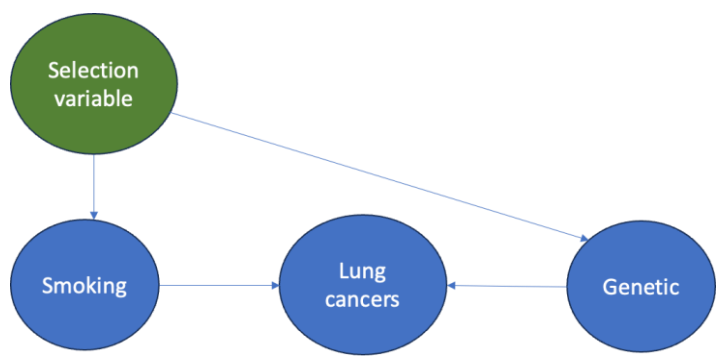

Figure 3: A selection diagram showing factors causing lung cancers

Building on the selection diagram, Subbaswamy and colleagues developed the surgery estimator. They employed the do-operator to eliminate the influence of selection variables within an interventional distribution, i.e. removing an arrow from 'Development or Target population' to 'Smoking' in figure 3[19]. The interventional distribution is the distribution under the assumptions that the variables (in this case the selection variables), had been set in a hypothetical scenario, which is possibly contrary to the observational scenario [41]. According to the previous selection diagram, the interventional distribution reflects the distribution of lung cancers, smoking, and genetic factors, if all individuals had been selected from the target population (which might be contrary to the fact that some individuals are from the development population). This removes the arrow from the selection variable to smoking. To make the interventional distribution more applicable with the observed datasets, Subbaswamy introduced the graph surgery estimator, which applies an identifiable algorithm to transform interventional distributions into observational ones. This transformed distribution, when mapped onto empirical data, facilitates predictions in hypothetical scenarios where selection variables are removed, enabling the transport of predictions to other populations. In essence, the surgery estimator is about starting with a selection diagram, in which we identify which causal relationships we believe will transport and which will not – through selection variables – then building models exploiting that assumption.

In addition to the selection diagram, Bellamy et al. proposed the characterisation of a shortcut predictor (feature) on a DAG [22]. A shortcut predictor is any predictor that is associated with an outcome and is site-specific. For example, a watermark in a hospital that specifically admitted COVID-19 patients. Identifying and removing shortcut predictors require expert knowledge on causal relationships of them and true predictors. Excluding the shortcut predictor before training a CPM is necessary for it to be generalisable. The shortcut predictor is essentially the same idea as the selection diagram previously described.

Minimising worse-case loss with mutable and immutable sets

This strategy leverages external knowledge to identify a set of predictors that adhere to the invariant property of the prediction system. This property is characterised by elements within the system that remain consistent, or invariant, across various populations, represented by immutable variables, and other elements that change across the system, represented by mutable variables. For example, when developing a CPM for prediction of cardiovascular disease using multi-centre data, we expect that the patient's risk factors such as age, gender or family history are similar – or immutable (if we assume that each centre represents the same distribution), however, the patterns of treatments such as statins prescription might be different – or mutable – due to each centre's policies. In this case the selection of variable should be directed toward immutable rather than mutable ones. In a paper by Subbaswamy et al., the authors proposed a method to identify the model that minimises the expected loss in the worst-case subpopulation, which is delineated by the sets of mutable and immutable variables [31]. The authors acknowledged that this method has the potential to guide the selection of generalisable prediction models that are stable to data distribution shift. However, they did not clearly explain how to achieve this. Further research could address this point. In addition, as a selection variable defines a variables that has structural difference in populations, which is a mutable variables, potential research could be done to explore how tackling generalisability and/or transportability by surgery estimator would achieve similar model performance as minimising worse-case loss with mutable and immutable sets does.

In addition, Liu (2023) adopted causal reasoning to distinguish between mutable and immutable variables [30]. Liu utilised minimax optimisation techniques based on these variable types and the specified causal graph to find an optimal subset of covariates that ensures stability, measured by the worst-case risk when transferring between populations. This subset serves as the starting point for developing generalisable and transportable models, however, similar to the previous methods from Subbaswamy et al., further research are needed on optimal methodology to develop a prediction model on this subset.

Causal predictor selection

Lemmon et al. applied predictor selection strategy motivated by causal structure[38]. This is done by identifying predictors that reside in the Markov Blanket (MB) and those that are parents or children (Parent Child - PB) of the outcome. Based on a DAG incorporating causal relationships among predictors and an outcome, the Markov Blanket (MB) is a subset of the DAG that shields the outcome variable from other predictors (figure 4). That is, when we know all values of predictors residing in the MB, we will have all the information to predict the outcome, regardless of other predictors. Lemmon termed selection criterion with MB or PB as causal predictor selection, which utilises causal knowledge to correctly choose variables which are MB or PB, thereby orienting it more towards a knowledge-driven approach. Similarly to the predictor selection methods, selecting predictors causally related to the outcome, either through MB or PB, contributes to improved generalisability by making the model more parsimonious. This approach also reduces the chance of selecting variables that vary across different settings. This strategy was also done by Piccininni (2020), who formalised the approach that, based on a prescribed causal graph, the subset of predictors needs to be chosen only from those that are parents (causes) of the outcome, not the children (consequences) (6). This is because the probability of a consequence given its cause is invariant across any populations, while the probability of a cause given its consequence is not. The scenario that is applicable in this context is that we are predicting disease outcomes based on their risk factors – prognostic prediction. However, this methodology is not useful in diagnostic prediction where the parents, i.e. disease of interest, are predicted based on the children, i.e. symptoms.

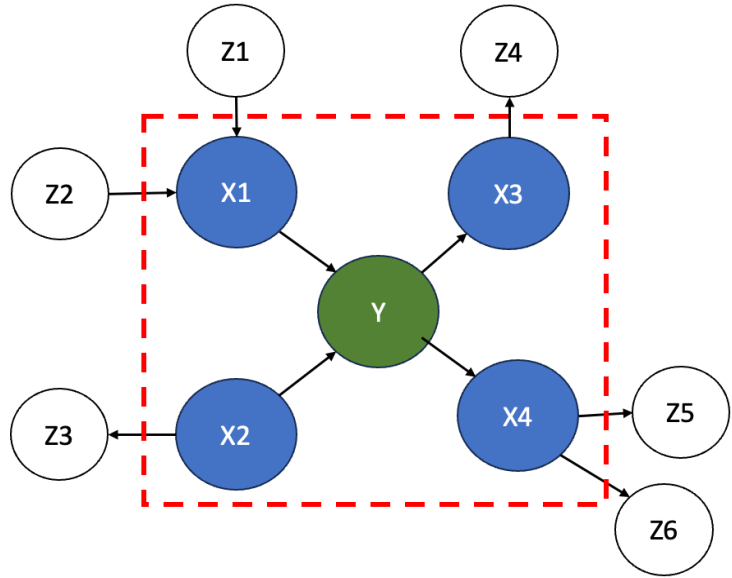

Figure 4: Prediction of Y with only variables that reside in Markov Blanket (red dash square).

**Knowledge-driven methodology that is tailored for a specific target population**

Neto et al. proposed a causality-aware prediction approach for anticausal predictions where predictors are caused by a predicted outcome, i.e. diagnostic prediction [36]. For example, it involves prediction of lung cancers having cough or chest radiography as predictors. This method aims to predict the outcome of interest by training the prediction model on the simulated counterfactual predictors including only those that are direct effects of the outcome. For example, the authors simulated counterfactual data containing only the associations generated by the direct effect of the outcome (which is the effect of interest). It is particularly important when confounding is unstable between development and target populations. Interestingly, they found that the models that were adjusted for confounding using only development data are not robust for prediction in the target data when dataset shifts generated by selection bias is present. To improve the model's robustness and thus the transportability to the target population, they suggested that the data from the target population is required to adjust for confounding on the target population. However, the authors examined the method on diagnostic prediction, which involve an anticausal direction. Causaility-aware prediction for prognostic model on causal direction where an outcome is predicted by its causes may not have the same behaviour as per [23].

**Discussion**

We have categorised approaches to aid the generalisability and transportability of clinical prediction models into two themes: one theme is data-driven, aimed at enhancing model robustness and only informed by empirical data. The other theme is knowledge-driven, which incorporates external knowledge in a form of causal relationships before using development data to train a clinical prediction model. The approaches can also be categorised according to whether they additionally take data from a target population during model development. Taking data from the target population as part of model development results in tailored models for that specific target population. Models that do not use data from any target population focus on generalisability to a range of possible target populations.

For the data-driven approach, the concept of developing generalisable prediction models by increasing heterogeneity during development (enhancing between-cluster heterogeneity) and validation (utilising internal-external validation) was also previously proposed by Debray et al.[4], however, their paper was not included in this review as it did not match our search keywords. Bareinboim and Pearl had previously established a formal definition for the transportability of experimental findings from heterogeneous development population regarding causal effects [42]. Connecting this work with their earlier work on observational transportability (definition 5 in[37]), which states that a statistical relation can be transported to a new domain by reconstructing it using data from the new domain, including only a subset of variables, there is an implication that this definition of observational transportability could be modified to formalise the transportability of predictions from a heterogeneous development population. This interpretation supports the proposals by both Debray and colleagues [4] and Van Calster et al. [43], and aligns with the findings of this scoping review (increasing heterogeneity of development population). Future research could significantly contribute to this area of development. For example, to improve the transportability of clinical prediction models constructed using stepwise internal-external validation (SEICV) [24] on heterogenous development population, the method can be integrated with $\mu$-Transportability [42]. $\mu$-Transportability defines the transportability of causal relations from heterogenous source domains. However, accessing heterogeneous data is not straightforward in real-world settings because of a variety of factors, including administrative, governance, privacy, or technical barriers. Large healthcare databases are a valuable repository of heterogenous data, yet they are not totally exempt from those concerns[44]. As an alternative, the ensemble methodology promotes diversity at model levels and can mitigate privacy and governance concerns since base learners can be trained locally, reducing the need to centralise patient data [27]. However, large healthcare databases could be employed with data anonymisation techniques or General Data Protection Regulation (GDPR) standards to improve patient's privacy so that these ensembles' advantages may be no longer be required.

In addition to increasing heterogeneity, we also found some papers suggesting that the generalisability of a clinical prediction model could be improved by ensuring the model contains only important predictors [38, 45]. We did not include these papers in the current review as there were insufficient theoretical information to support how this method would improve generalisability and/or transportability. The hypothesis is that such model is parsimonious and would include only predictors that are invariant to the data distribution shift, thereby improving the generalisability and/or transportability[38]. These methods include 1) Regularisation [38], 2) Remove and Retrain (ROAR) [38] and 3) Shapley value [45]. Regularisation shrinks the predictor coefficients toward zero. It reduces model variance and hence overfitting to development data. However, there is no guarantee that taking shrunken coefficients fitted from development population would improve generalisability and transportability in target population. ROAR and Shapley value work similarly in that they identify important predictors based on a permutation-based importance score (ROAR), or how each predictor performs on prediction task against the average prediction task (Shapley). According to the original papers, regularisation and important predictor selection yielded generalisable prediction in some circumstances, however, it is still not clear how the methods improve generalisability or transportability.

Turning to the knowledge-driven approach, it might initially seem to be a natural method for training generalisable or transportable prediction models, as it aims to capture and learn from causal structures that should theoretically aid generalisability and transportability. However, the process is not as straightforward as it appears. As the approach's name implies, it entails leveraging domain knowledge of the underlying causal structure to construct structural causal models or graphs. This method is somewhat equivalent to estimating causal effect using randomised or observational studies, in such a way that it leverages domain knowledge to construct the underlying causal structure of the problems being studied. Though both requires assuming that the causal models are true, the causal effect estimation by randomisation have fewer causal assumptions, particularly on the study population. A common thread among the knowledge-driven methodologies reviewed in this study is the reliance on strong assumptions regarding accurately specified causal structures. This is particularly true for the specifications of selection variables or assumptions about specific covariates causing variations in the distributions between development and target populations. To best of our knowledge, no studies have yet proposed a methodology that estimates causal effects that is resistant to effect modifiers and use these effects to construct generalisable prediction models with minimal extra assumptions. There is a growing literature proposing methods to transport a causal effect to the target population in the presence of effect modifiers [46–48]. Lin et al. [49] identified causal methods that allow predictions under hypothetical interventions. We could view these interventions as situations in which an individual who is actually in the development population would have been reassigned to the target population. This could possibly answer generalisability and transportability problems. Identifying causal methods to enable prediction under hypothetical interventions of "switching" the populations could be the next direction of methodological research on generalisability/transportability.

Addressing generalisability and transportability of CPMs from a data-driven perspective is not straightforward. Traditionally, developed models are externally validated in different populations and settings to ensure that they would perform as they did in development settings. Van Calster et al. argue that an external validation only captures generalisability and transportability on a single external dataset so says nothing about 'generalisability in general' [43]. However, the distribution of the population underlying the evaluating external dataset can change from the point at which the external validation is taken, to the point at which the model is being implemented, there is no reason to assume that models that have been evaluated with the external validation are always transportable. This is because though the models are perfectly validated in the static external validation, there is no guarantee that the models would be well performed in the new, changing environments [43]. Instead, they offer a series of recommendations for model development and validation that align with the data-driven methodology presented in this scoping review, which includes incorporating greater heterogeneity in the training and validation processes.

Recent studies have conducted methodological reviews on generalisability and transportability from various perspectives: one addresses the external validity of causal effects, including heterogeneity of treatment effect in terms of generalisability and transportability[50], while another focuses on generalising or transporting treatment effects from randomised clinical trials to target populations[51]. Furthermore, there exist the comprehensive surveys on transfer learning[52, 53], a subfield in machine learning that aims to enhance the learning of prediction models in the target population by utilising knowledge derived from models previously trained in the development population. This facilitates the transfer of knowledge from development to target. Although transfer learning and transportability are not equivalent, their methodologies share many similar characteristics that reveal some overlap between these domains. Accordingly, there are studies that have adopted transfer learning approaches, such as density ratio weighting, to improve the transportability of clinical prediction models or causal effects to a target population[26, 28, 32, 54–58]. While this current review did not include these methodologies, which were uncaptured with the search terms, it contributes to the body of knowledge on generalisability and transportability within predictive modelling by demonstrating that the methodologies aiding generalisability and transportability of clinical prediction models have been proposed from various fields of study. Previously, there are reviews on model updating, which is also essentially a form of improving the transportability of a CPM [11, 12].

This review has some limitations. As the studies it encompasses originated from two distinct communities: medical statistics and machine learning. Consequently, the terminologies employed to articulate the act of generalising or transporting findings from a development to a target population differ. This diversity presents a challenge in designing search terms that are comprehensive enough to capture relevant articles while excluding non-pertinent ones and avoiding the omission of seminal works. Martin et al.[21] addressed this issue by proposing an iterative search-and-refine strategy to optimally identify pertinent literature. Although this review has adhered to such an approach, the notable variation in language among both disciplines [59], raises the possibility that some relevant papers may have been missed. This became evident when one of the five known papers was not captured by the search terms and remained undetected through citation searching. Moreover, most exclusions took place during the abstract screening phase, notably omitting studies concerning the external validation of clinical prediction models. Abstracts from such studies frequently have insufficient detail to determine whether the methodologies employed were specifically designed to address transportability issues. Consequently, this could result in the inadvertent exclusion of relevant methodologies that only apparent upon full text review. This poses a significant challenge when screening a large volume of papers based solely on abstracts. Another limitation of the search strategy is that certain databases, such as Scopus or Web-of-Science were not included, which may have led to the missing of relevant studies, particularly those outside the medical field.

This scoping review identifies several key gaps for future research to develop optimal methodology aiding generalisability and transportability of clinical prediction models. Firstly, we can design an observational study that mimics a randomised trial (e.g. target trial emulation, [60]) to estimate the causal effect of being in the different populations on the outcome of interest. For example, to facilitate transportability of cardiovascular disease (CVD) prediction developed from secondary care population to general healthy population, we can estimate the effect of belonging to a healthy population that has never visited GPs, instead of being hospitalised patients, on the risk of the disease. This effect estimate could be further embedded in generalisable or transportable prediction models. Secondly, methodologies that integrate data-driven techniques—like the use of heterogeneous data—with robust causal approaches could be particularly useful. For example, the application of counterfactual predictions from observational data may be a step in this direction. Thirdly, as several methodologies necessitate access to target population data, which have the strengths of directly tailoring a prediction model to the target population, there is a need for methods that reduce this requirement but still retains the ability for the transportability to the population of interest. This may include creating hypothetical population where the distribution is assumed to be similar as the target population. Finally, there is a need to evaluate the performance of these proposed methodologies on both simulated and real-world datasets that present various types of dataset shifts, to understand their efficacy in broader contexts.

## Conclusion

In conclusion, this scoping review suggests that methodology to aid generalisability and transportability of clinical prediction models can be categorised as data-driven and knowledge-driven. Data-driven methods focus on harnessing actual data to maximise model robustness, including heterogeneity or importance weighting. In contrast, knowledge-based methods utilise causal reasoning to form underlying invariant structure that encoding generalisable or transportable prediction models. Future work could explore integrating both approaches into a single, unified one for better generalisability and transportability

## Statements and declarations

## Ethical considerations

This article does not contain any studies with human or animal participants.

## Consent to participate

Not applicable

**Consent for publication**

Not applicable

**Declaration of conflicting interest**

The authors declared no potential conflicts of interest with respect to the research, authorship, and/or publication of this article.

**Data availability**

Data sharing not applicable to this article as no datasets were generated or analysed during the current study.

Supplemental table 1: Summary of included 18 papers

| No | Authors | Methodology | Description | Need target population? | Domain knowledge? | Stated assumptions | Reported advantages | Reported limitations | Experiments |
|---|---|---|---|---|---|---|---|---|---|
| 1 | Gao et al. [26] | (D) Density ratio weighing | These are the methods that were used to estimated the density ratio:<br>1. Direct methods<br>- KMM and RuLSIF<br>2. Indirect methods<br>- LR with L1 and L2 penalty, and Random Forests | Yes, note that outcome data is required only for the indirect methods | No | The distribution shift is compatible with covariate shift | Estimation of density ratio with both parametric and non-parametric models are useful for covariate shift correction. | Using density ratio weighting is not possible to correct concept shift. | - Applied the methodology on real-world eICU data<br>- Simulated dataset for prior shift and concept shift |
| 2 | Liu et al. [30] | (D) Predictor selection of invariant information from a causal perspective | The authors propose that selecting the entire set of stable predictors, S, is not always optimal, as retaining all stable information may not yield the best results. This is determined by a graphical condition based on the causal relationships encoded in the data, which can be verified through causal discovery methods. The condition specifically assesses whether the outcome variable significantly impacts a subset of mutable variables that are descendants of another set directly related to the target. If S is not deemed optimal, the authors introduce an optimisation algorithm to identify a subset of predictors with the minimal worst-case risk. This algorithm starts by learning a causal graph from the training data, encompassing the outcome and predictors. It then checks the influence of the outcome variable on the specified subset of variables to decide if the full set S is optimal. If not, the algorithm categorises predictors based on their risk profiles and selects the group with the lowest risk, ensuring the model's robustness across varying data conditions. | No | Yes | The methodology is based on a faithful structural causal framework and assumes a Markovian structure. It assumes that the distributional shifts are only due to changes in the mutable variables' distribution. Though, correctly identifying them is particularly difficult when working with real-world data,. | It introduces a novel optimisation algorithm to estimate the worst-case risk for each subset of stable predictors and selects the one with the minimal risk. | The effectiveness of the methodology is depend on the accuracy of the assumed causal framework, including the specification of mutable and immutable variables. In addition, it requires computational resources for causal discovery and the evaluation of worst-case risks. | The authors tested their causal minimax learning approach against various baselines such as Vanilla, ICP, IC, DRO, and others. It used metrics including max MSE and std MSE for evaluation. The synthetic study involved data generation across 20 different environments using a particular DAG and structural equation setup, where the distribution of a mutable variable varied. The authors also evaluated a predictive model for Alzheimer's disease, employing neuroimaging data to identify key predictors, with distinct environmental patterns observed across various age groups. |

| No | Authors | Methodology | Description | Need target population? | Domain knowledge? | Stated assumptions | Reported advantages | Reported limitations | Experiments |
|---|---|---|---|---|---|---|---|---|---|
| 3 | Subbaswamy et al. [31] | (D) Robustness and stability assessment | - Classifying predictors: Classifying predictors into immutable and mutable sets.<br>- Defining distribution shifts: Factoring the probability distribution to focus on the impact of changes in mutable variables' distributions.<br>- Assessing Robustness: Evaluating worst-case risk by identifying subpopulations with the worst average loss under distribution shifts. | No | Yes, to define mutable/immutable variables and formulate distribution shift | - The subpopulation identified for estimation of worst-case performance is representative of potential real-world shift<br>- The mutable, immutable variables and their distributions are assumed correctly. | - Allow for estimate performance in worst-case populations, reflecting stability, without need to collect target data | Not stated | - Evaluated classical and robust models, for predicting sepsis with 17 predictors using EHR data from hospital A over four years<br>- Compared estimated worst-case performance under difference in pattern of lab test ordering |
| 4 | Piccininni et al. [23] | (D, K) Selecting predictors which is a parent (cause), not a child (consequences) | The principle of independent mechanism states that P(cause), and P(effect|cause) contain no information about each other, both are independent of each other but not the anti-direction. This independence makes effect prediction model with P(efffect|cause) stable and transportable to another settings where P(effect) changes, i.e. covariate data shift. | No | Yes, to build a faithful DAG | DAG must be correctly specified (faithfulness) and contains no hidden variables. | The model is simple to understand. | DAG must be correctly specified (faithfulness) and contains no hidden variables. | There are no experiment involving how prediction selection with MB would impact transportability (the authors did simulated experiments, but mainly focused on internal validation of CPM with various predictor selection strategies). |

| No | Authors | Methodology | Description | Need target population? | Domain knowledge? | Stated assumptions | Reported advantages | Reported limitations | Experiments |
|---|---|---|---|---|---|---|---|---|---|
| 5 | Steingrimsson et al. [32] | (D) Importance weighting by density ratio estimation | Correctly-Specified Model: Tailoring the model for the target population is not required.<br><br>Misspecified Model: The tailoring involves estimating the density ratio, which is done by using the inverse of the odds of being from the development population as weights. These weights are then applied in the prediction model to adjust it for use in the target population. This step is crucial to account for differences in the distribution of covariates between the development and target populations. | Yes, covariate data | Yes | 1. Conditional independence of the outcome and the population: this requires that the outcome is independent of selection in the population conditional on covariates, which implies conditional mean exchangeability over the population. According to this assumption, the relationship between the outcome and covariates remains constant common across populations.<br>2. Positivity: this condition necessitates that for every covariate pattern present in the target population, there is a positive probability of encountering the same pattern in the development population. | The methodology aids tailoring a prediction model from the development population to the target population by employing weighted adjustments to individual data points without the need for re-training the model. Additionally, these weighted approaches provide insights into the variations in the model's performance between the development and target populations. | 1. The applicability of the methods may be limited if there is a lack of access to covariate data from the target population.<br>2. The set of covariates necessary to satisfy the conditional independence assumption might be extensive than the set actually used in the model.<br>3. The approach did not address the practical issues such as missing data and measurement error. | The authors used simulated data to test the performance of the prediction models. Data was generated using a linear model with heteroscedastic errors and simulated membership for development population using logistic regression. They assessed both with Ordinary Least Squares (OLS) and Weighted Least Squares (WLS) regresson. WLS incorporating weights based on inverse odds of being in the development data training sets.<br>They also used real-world data for prediction of lung cancer diagnosis in the US using data from the NLST and the NHANES. A prediction model was tailored with demographic and health predictors using inverse-odds weighted logistic regression to transport a the trial (NLST) to a national setting (NHANES). |
| 6 | Schinkel et al. [33] | (D) Induce training data heterogeneity | AI models trained on multiple cohorts with heterogeneity and diluted hospital-specific patterns, are more generalisable. | No | No | Heterogeneous cohort data is available. | - | - | Testing patients having blood cultures in ER in different medical centres to predict poor outcomes |

| No | Authors | Methodology | Description | Need target population? | Domain knowledge? | Stated assumptions | Reported advantages | Reported limitations | Experiments |
|---|---|---|---|---|---|---|---|---|---|
| 7 | Gong et al. [34] | (D) Instance weighting | - Weighted Instances: Calculates Euclidean distance of each instance (development and target) from target mean; inverts these distances against the maximum target distance to assign weights.<br>- Weighted Clusters: Uses K-means to cluster development data, then measures Euclidean distance between cluster means and target mean for weighting; closer clusters get higher weights. | Yes | No | - Assumes dataset shift as covariate shift.<br>- Transfers insight from numerous cardiac surgeries.<br>- Development data suitable for clustering in Weighted Clusters method. | The methods perform well in small datasets. | The improved performances were not statistically significant, so the methods are needed to be tested on larger multi-center datasets. | - Analysing data from 2 hospitals, (2001-11), for cardiac surgery risk stratification.<br>- Preprocesing this data to include about 250 features, employing (CMIM) to avoid overfitting.<br>- Fitting models with L2-regularised logistic regression, to address class imbalance<br>- Taking evaluation through LOOCV, measuring performance with precision, recall, and AUC metrics |
| 8 | Lemmon et al. [38] | (D) Predictor selection strategies that enable robustness to temporal dataset shift | Causal feature selection: select features based on causal relationship with the outcomes, selected predictors were in Markov Blanket (MB) or Parent-Child (PB)<br>In addition, there were two proposed methods which we were not include in our review due to insufficient information:<br>1. L1-regularisation<br>2. Remove and Retrain (ROAR): identified importance predictors with a permutation-based feature importance score and removed, then retrained the model | No (requires only historical data) | No (except for causal predictor selection) | No explicitly mentioned<br>- it is inferred that the assumption for temporal shift relates to the invariant relationship between outcome and predictor - i.e. covariate shift | Assessing multiple feature selection methods for transportability across several clinical outcomes | - Only used MIMIC-IV dataset for evaluation.<br>- Inability to predict future predictor shifts.<br>- Lack of a - definitive MB ground truth in MIMIC-IV.<br>- Incomplete assessment of alternative methodologies. | - Analyzed ICU patient data from the MIMIC-IV database.<br>- Built logistic regression models trained on 2008-2010 data to predict long LOS, in-hospital mortality, invasive ventilation, and sepsis for 2011-2019.<br>- Evaluation metrics used: AUROC, AUPRC, ACE. |

| No | Authors | Methodology | Description | Need target population? | Domain knowledge? | Stated assumptions | Reported advantages | Reported limitations | Experiments |
|---|---|---|---|---|---|---|---|---|---|
| 9 | Reps et al. [27] | (D) Model ensemble | The authors proposed binary ensemble classifiers to increase the transportability of prediction models. These ensembles are formed by combining individual LASSO logistic regression classifiers, each trained on different databases but predicting the same outcome. The weighted fusion ensemble method assigns weights to these base classifiers, based on their performance metrics or the similarities between the populations they were trained on. It aims to optimise the overall accuracy by effectively utilising the strengths of each classifier. The stacking ensemble method uses predictions from these base learners as input features for a secondary classifier. This secondary classifier adaptively learns the optimal weighting for each base learner's prediction, thereby improving the ensemble's accuracy by tailoring it to the unique characteristics of the dataset being applied to. | Yes | No | Did not explicitly mention | The study highlights the strength of the ensemble methods in predicting outcomes for new populations, because of their ability to integrate multiple models trained on diverse datasets. It also was able to compare the transportability of base learners and ensembles in different databases. | The limitations of this study is the generalisability of findings, because the authors studied only for one target population in the US. Miscalibration was also observed for all models. | The study was designed to predict various poor outcomes in patients with pharmaceutically-treated depression, utilising data from both US claims and Electronic Health Records (EHR). The data comprised five datasets, each representing patients with varying characteristics. For model development, base learners were trained using LASSO-penalised logistic regression. To assess the transportability across databases, a leave-one-database-out approach was employed. In this approach, each of the five datasets was excluded in turn to evaluate the performance of the ensemble models. The ensemble models were trained on the remaining four databases and then tested on the excluded database to assess their discrimination and calibration metrics. This process was repeated for each of the five datasets. |
| 10 | de Jong et al. [24] | (D) Predictor selection using Internal-External Cross-Validation (IECV) of heterogenous training data | 1. Clustering the training dataset to mimick real-world heterogeneity and use ICEV for those clusters to perform stepwise predictor selection  2. From stepwise IECV, aggregate performance metrics of all clusters using meta-analytic methods to select the best model that can generalise across heterogenous dataset | No | 1. The choices of candidate predictors should be guided by clinical knowledge. | Heterogeneity of training data | - | 1. Predictors are not related to heterogeneity in predictor-outcome associations.  2. Generalisability may come at a cost of improving one predictive performance (such as discrimination), but deteriorating another one.  3. Focusing solely on predictor selection may have an impact on model's precision.  4. It is still possible that the model requires recalibration on the target settings. | In their prior work, the author developed site-specific XGBoost models for three healthcare systems (KPSC, IH, UWM) using over 200 candidate features. The top 20 features, each with importance values ≥1%, were identified as most predictive. Simplified models using these top features were built for each site. While these simplified models performed well within their respective sites, their performance decreased by up to 4.1% when applied to other sites |

| No | Authors | Methodology | Description | Need target population? | Domain knowledge? | Stated assumptions | Reported advantages | Reported limitations | Experiments |
|---|---|---|---|---|---|---|---|---|---|
| 11 | Dockès et al. [28] | (D) Importance weighting of training data | Use density ratio weighting to adjust covariate balance between development and target populations | No | No | Outcome is independent of Selection given covariates | - | - | - |
| 12 | Pajouheshnia et al. [29] | (D) Comparability of measurement methods | The discriminative ability of a prediction model varies across settings with variation in the measurement of predictors | Yes | Yes | Assuming normal distribution of predictors | First paper to consider measurement errors as transportability issues | Isolate the effect of predictor measurement from other factors | Case study in DVT prediction to show how increasing measurement errors in different settings affect discrimination performance |
| 13 | Qian et al. [35] | (D) Adversarial validation | Adversarial validation by mimicking validation data with target population: | Yes | No | - | - | - | Credit score prediction |
| 14 | Subbaswamy et al. [19] | (K) Using causal methods to help transporting a CPM encoded by a graph | Proposing the Graph Surgery Estimator (GSM) by creating a selection diagram having the interventional distribution on the mutable variables and then apply the identifiable algorithm to resolve into the observational distribution | No | Yes, domain knowledge require for both causal relationships (including unobserved confounders) of each variables in the system, and specification of the variables that vary across enviroments (mutable variables) | 1. The selection diagram is correctly specified. 2. The interventinal distribution is identifiable. | | 1. Predicting variables is mutable. (prior probability shift) | 1. Simulating data from DGP by pre-defined DAG and generate different environments by vary mutable variables. 2. Using UCI Bike sharing dataset to predict the number of hourly bike rentals in different seasons by assuming possible DGP as a DAG including mutable variables. |

| No | Authors | Methodology | Description | Need target population? | Domain knowledge? | Stated assumptions | Reported advantages | Reported limitations | Experiments |
|---|---|---|---|---|---|---|---|---|---|
| 15 | Pearl et al. [37] | (K) Formalise transport formula using selection diagram | The statistical relation of interest, such as P(y\|x), can be estimated in the target domain using the following transport formula: P*(y\|x) = ∑c P(y\|x,C) P*(C), where P*(y\|x) represents the transported relation in the target domain, C is a set of covariates that satisfies the condition of S-admissibility. The key aspect is that the marginal probability distribution of C, P*(C), is the only one that need to be re-estimated in the target domain. | Yes | Yes | 1. Valid selection diagram is required 2. The covariate C need to satisfy S-adminissibilty, i.e. y ⊥ x \| C | | - Mainly addresses the causal effect of a single variable X and does not cover prediction scenarios with multiple predictors<br>- Lacks cases where the covariate set C overlaps with the predictor set X, which is common in observational studies<br>- Predominantly remains theoretical | - |
| 16 | Fehr et al. [20] | (K) Using DAG to help transporting a CPM | Model that takes causes as predictors is more transportable than those taking consequences | No | Yes, to build a DAG | Relies on underlying structural causal equations. | Novel approach to assessing the transportability of CPMs using empirical data and underlying causal structures | DAG needs to be verified. | Using ADNI data to create DAGs and use structural equations to quantify the relationship for the equations, then use the DAGs to test transportability under different situations |
| 17 | Neto et al. [36] | (K) Counterfactual approach for causality-aware predictions | The methodology aims to adjust confounding in anticausal prediction tasks. The goal is to generate causality-aware predictors that account for confounding variables. Using predictors, confounders and outcomes from training data, the methodology estimates the regression coefficients and residuals and uses this information to create a counterfactual version of predictors that adjusted to remove the influence of confounders. Apply the same process to the test or target data to adjust using the coefficients and residuals from the training set to create the logically-similiar counterfactual set of target data. | Yes, covariate data | Yes | The assumptions need correctly specifying that the predictors are caused by an outcome - anticausal prediction. In addition, working with causal relationship requires domain knowledge. | It does not require full understanding of causal graph, only the causal relationships of confounders, predictors and outcomes. | The methodology focuses predominately on linear models. It also sensitive to inaccurate specification of variables in the system. | The authors employed linear models with ten predictors and one confounder, generating samples of training and test sets. They imposed a covariate shift to mimic changes in the joint distribution of confounders and the outcome that resulted from selection biases. The design aimed to reveal that only adjusting confounders in training data is not enough for robust predictions in the face of dataset shifts caused by selection bias. |

| No | Authors | Methodology | Description | Need target population? | Domain knowledge? | Stated assumptions | Reported advantages | Reported limitations | Experiments |
|---|---|---|---|---|---|---|---|---|---|
| 18 | Bellamy et al. [22] | (K) Shortcut predictor | A shortcut predictor is any predictor that is associated with an outcome and is site-specific. For example, a watermark in a hospital that specifically admitted COVID19 patients. Identifying and removing shortcut predictors require expert knowledge on causal relationships of them and true predictors. Excluding the shortcut predictor before training a CPM is necessary for it to be generalisable. | No | Yes | Correctly specified DAG | - | - | - |